\documentclass[conference]{IEEEtran}

  	\usepackage[pdftex]{graphicx}
  	\graphicspath{{../pdf/}{../jpeg/}}
	\DeclareGraphicsExtensions{.pdf,.jpeg,.png}

	\usepackage[cmex10]{amsmath}
	\usepackage{mathabx}
	\usepackage{algorithmic}
	\usepackage{array}
	\usepackage{mdwmath}
	\usepackage{mdwtab}
	\usepackage{eqparbox}
	\usepackage{url}
	\graphicspath{ {images/} }

\usepackage[colorinlistoftodos]{todonotes}

\usepackage[colorinlistoftodos]{todonotes}

\usepackage[colorinlistoftodos]{todonotes}

\usepackage[colorinlistoftodos]{todonotes}

\begin{document}

\title{
Control of the phase of reflected spin-waves from magnonic Gires-Tournois interferometer of subwavelength width

}

\author{\IEEEauthorblockN{1\textsuperscript{st} Krzysztof Sobucki}
\IEEEauthorblockA{\textit{Institute of Spintronics}\\
\textit{and Quantum Information,}\\
\textit{Faculty of Physics,}\\
\textit{Adam Mickiewicz University,}\\
\textit{Uniwersytetu Poznańskiego 2,} \\
\textit{Poznan, Poland}\\
krzsob@amu.edu.pl}
\and
\IEEEauthorblockN{2\textsuperscript{nd} Paweł Gruszecki}
\IEEEauthorblockA{\textit{Institute of Spintronics}\\
\textit{and Quantum Information,}\\
\textit{Faculty of Physics,}\\
\textit{Adam Mickiewicz University,}\\
\textit{Uniwersytetu Poznańskiego 2,} \\
\textit{Poznan, Poland}}
\and
\IEEEauthorblockN{3\textsuperscript{rd} Justyna Rychły}
\IEEEauthorblockA{\textit{Polish Academy of Sciences} \\
\textit{Institute of Molecular Physics}\\
Poznań, Poland }
\and
\IEEEauthorblockN{4\textsuperscript{th} Maciej Krawczyk}
\IEEEauthorblockA{\textit{Institute of Spintronics}\\
\textit{and Quantum Information,}\\
\textit{Faculty of Physics,}\\
\textit{Adam Mickiewicz University,}\\
\textit{Uniwersytetu Poznańskiego 2,} \\
\textit{Poznan, Poland}}
}

\maketitle

\begin{abstract}

The phase is one of the fundamental properties of a wave that allows to control interference effects and can be used to efficiently encode information. We examine numerically a magnonic resonator of the Gires-Tournois interferometer type, which enables the control of the phase of spin waves reflected from the edges of the ferromagnetic film. The considered interferometer consists of a Py thin film and a thin, narrow Py stripe placed above its edge, both coupled magnetostatically. We show that the resonances and the phase of the reflected spin waves are sensitive for a variation of the geometrical parameters of this bi-layerd part of the system. The high sensitivity to film, stripe, and non-magnetic spacer thicknesses, offers a prospect for developing magnonic metasurfaces and sensors. 
  
\end{abstract}


\section{Introduction}

Controlling the phase of waves, regardless of their type, is very important from the application's point of view. Among other things, phase control enables to encode information, e.g., in phase-shift keying digital modulation scheme widely used in modern wireless communication\cite{blahut1987InformationTheory}, while a wave-front modulation makes it possible to control the direction of propagation and a wave focusing. Furthermore, by controlling the phase of the reflected waves, chromatic dispersion can be generated, which found application in laser pulse compression\cite{kuhl1986compression}. A commonly used optical system for this purpose is the Gires-Tournois interferometer (GTI)\cite{Gires1964}. GTIs are  Fabry-Perot interferometers operating in a reflection mode rather than transmission mode. In practice, a GTI can be made of two mirrors, the first partially reflecting and the second completely reflecting the incident radiation. For wavelengths close to those satisfying the Fabry-Perot resonance condition, a strong phase dependence of the reflected waves on the wavelength of the incident radiation appears. 

The subject of controlling the phase and amplitude of electromagnetic waves by two-dimensional surfaces composed of resonators of sub-wavelength dimensions has become central to the development of the concept of metasurfaces for electromagnetic waves\cite{Yu2014,yu2011light, Kumer2012,vashistha2017colorFilters}. 
This has contributed to the rapid development of photonics in recent years.

The concept of metasurfaces has also recently received attention in magnonics\cite{krawczyk14rev, chumak2015rev,chumak2019fundamentals}, which is a subfield of magnetism focused on spin waves (SWs), particularly in the context of their applications to information transfer and signal processing. The pioneering theoretical demonstration of metasurfaces acting on SWs enabled their focusing by applying modulation of the exchange interaction alongside the interface between two interconnected ferromagnetic layers\cite{Zelent19}. In the later studies, it was shown that by continuous varying the value of the magnetocrystalline anisotropy or saturation magnetization in a narrow region between the ferromagnetic waveguide and the thin film, an anomalous refractive effect can be obtained for SWs, allowing SWs to be efficiently bent in the waveguides\cite{Mieszczak2020}. This closely connects the magnonic metasurface studies with the graded-index approach in the design of magnonic devices\cite{Whitehead2018, whitehead2019graded}.  

Several reports are showing that the placement of ferromagnetic stripes over a ferromagnetic film or waveguide can be used to emit SWs or affect the amplitude and phase of SWs passing below that stripe\cite{kruglyak2017graded, au2012phaseshifter, au2012transducer, al2008evidence, zhang2019resonator,Yu2019,Che20}.
Recently, we have demonstrated with micromagnetic simulations that the phase of reflected SWs can be controlled by using a resonator with a sub-wavelength width placed over the edge of a thin film\cite{sobucki2021resonant}. In that study, a resonator made of a material with a lower saturation magnetization value than the thin film was considered, which allowed obtaining an additional pair of short-wavelength modes in the bi-layered region, being crucial for effective phase modulation. This was the demonstration of a GTI operating on SWs\cite{sobucki2021resonant}. In this work, we continue the idea of the magnonic GTI and analyze the system suitable for the experimental realization. We study the interferometer composed of permalloy (Py) and analyze how the thickness of the film and resonator, the width of the resonator, and the separation between them, affect the phase modulation of the reflected SWs.

The manuscript is organized as follows. Section \ref{Methods} contains the definition of equations used in the calculations and an explanation of the post-processing of raw data obtained in numerical simulations. In section \ref{Results}  the geometry of the system under consideration is presented and there is an extensive description of the results obtained from our calculations. Section \ref{Conclusions} includes the final conclusions of our results.

\section{Methods\label{Methods}}

    SWs dynamics in the ferromagnetic layers can be studied in a frame of the linearized Landau-Lifshitz equation with damping neglected, coupled with the Gauss law for magnetism. In the case of SWs propagating along the $x$-axis, being perpendicular to the orientation direction of the static magnetization (aligned along the $y$-axis) in in-plane, uniformly magnetized layers, these equations read as:
    \begin{subequations}
    \label{eq:linearised-equations}
        \begin{align}
          \partial_x \psi - \mathbf{\nabla} \cdot \left(\frac{l^2}{M_{\rm{S}}} \mathbf{\nabla} m_x\right) + \frac{H_{\rm{0}}}{M_{\rm{S}}} m_x - \frac{i \omega}{|\gamma| \mu_0 M_{\rm{S}}} m_z = 0, \\
          \partial_z \psi - \mathbf{\nabla} \cdot \left(\frac{l^2}{M_{\rm{S}}} \mathbf{\nabla} m_z\right) + \frac{H_{\rm{0}}}{M_{\rm{S}}} m_z + \frac{i \omega}{|\gamma| \mu_0 M_{\rm{S}}} m_x = 0, \\  
          \partial_x(m_x - \partial_x \psi) + \partial_z(m_z - \partial_z \psi) = 0 . 
        \end{align}
    \end{subequations}
    Solving these equations allows us to find the values of the dynamical components  of magnetization -- $m_x$ and $m_z$, and the magnetostatic potential -- $\psi$ of the eigenmodes, as well as the corresponding to them angular frequency $\omega = 2 \pi f$. Coefficients in Eqs.~(\ref{eq:linearised-equations}) are respectively: ${H}_0$ -- the static, uniform external magnetic field directed along $y$-axis, $M_{\mathrm{S}}$ -- the saturation magnetization of the ferromagnetic material, $\gamma$ -- the gyromagnetic ratio, $\mu_{0}$ -- the vacuum permeability, $l = \sqrt{ 2A_{\rm{ex}} / (\mu_0 M^{2}_{\rm{S}})}$ -- the exchange length, where $A_\mathrm{ex}$ is the exchange constant. 
    The terms including magnetostatic potential represent the influence of dipolar interactions, the terms with the exchange length -- the exchange interactions. Further method details and the approximations used can be found in Refs.~\cite{rychly2017spin, rychly2016spin}. 
    
    To solve the Eqs.~(\ref{eq:linearised-equations}), we use the frequency-domain finite element method (FD-FEM) defined in COMSOL Multiphysics software\cite{comsol}, as described in Ref.~\cite{rychly2017spin}. At the edges of the computational domain (far from the ferromagnetic materials) the Dirichlet's boundary conditions, forcing the magnetostatic potential to vanish, are imposed.

    The phase of  SWs reflected from the interface located at $x=x_0$ can be extracted from the SW mode profiles (eigenmodes) calculated by FD-FEM. These mode profiles correspond to the steady-states and thus the phase of reflected SWs can be calculated fitting the $m_x$ component of magnetization to the equation:
    \begin{equation}
        \begin{split}
            m_x(t; x) = a(t) \cos[k_x(x-x_0)-\varphi/2],
        \end{split}
        \label{eq:fit}
    \end{equation}
    where $a(t)$ is a wave amplitude coefficient independent from the position and $\varphi$ is the phase shift of the reflected wave. Further details of postprocessing can be found in Ref.~\cite{sobucki2021resonant}
    
    Noteworthy, to numerically calculate the dispersion relations, and therefore, to obtain explicitly eigenfrequency dependence on $k_x$ (the 
    wave vector of SWs propagating along the $x$-axis), we have implemented the Bloch boundary conditions at the lateral edges of an elementary cell of width $w=30$ nm: 
            $m_x(x,z)=\tilde{m}_{x}(x,z) e^{i k_x x}$, 
            $m_z(x,z)=\tilde{m}_{z}(x,z) e^{i k_x x}$, 
            $\psi(x,z)=\tilde\psi(x,z) e^{i k_x x}$,
 where the functions with tilde on the right-hand side of equations are periodic functions of $x$, with the period $w$. 
 
The dispersion relation of SWs propagating perpendicularly to the direction of the effective magnetic field in a single in-plane, uniformly magnetized infinite thin film can be also described analytically\cite{stancil2009spin}: 
\begin{equation}
    \omega^2 = \omega_0  \left(  \omega_0 + \omega_\mathrm{M} \right) + \frac{\omega_\mathrm{M}^2}{4}\left[1 + \mathrm{e}^{-2k_{x}d} \right],
    \label{eq:dispersion}
\end{equation}
where $d$ is the film thickness, 
$\omega_0=|\gamma| \mu_0 (H_0 + M_\mathrm{S}l^2 k_{x}^2)$, and $\omega_\mathrm{M}=|\gamma| \mu_0 M_\mathrm{S}$, see Ref.~\cite[Chapter 7.1]{mag_osci_wav}.
    

\section{Results\label{Results}}
   \begin{figure}[t!]
        \centering
        \includegraphics[width=7cm]{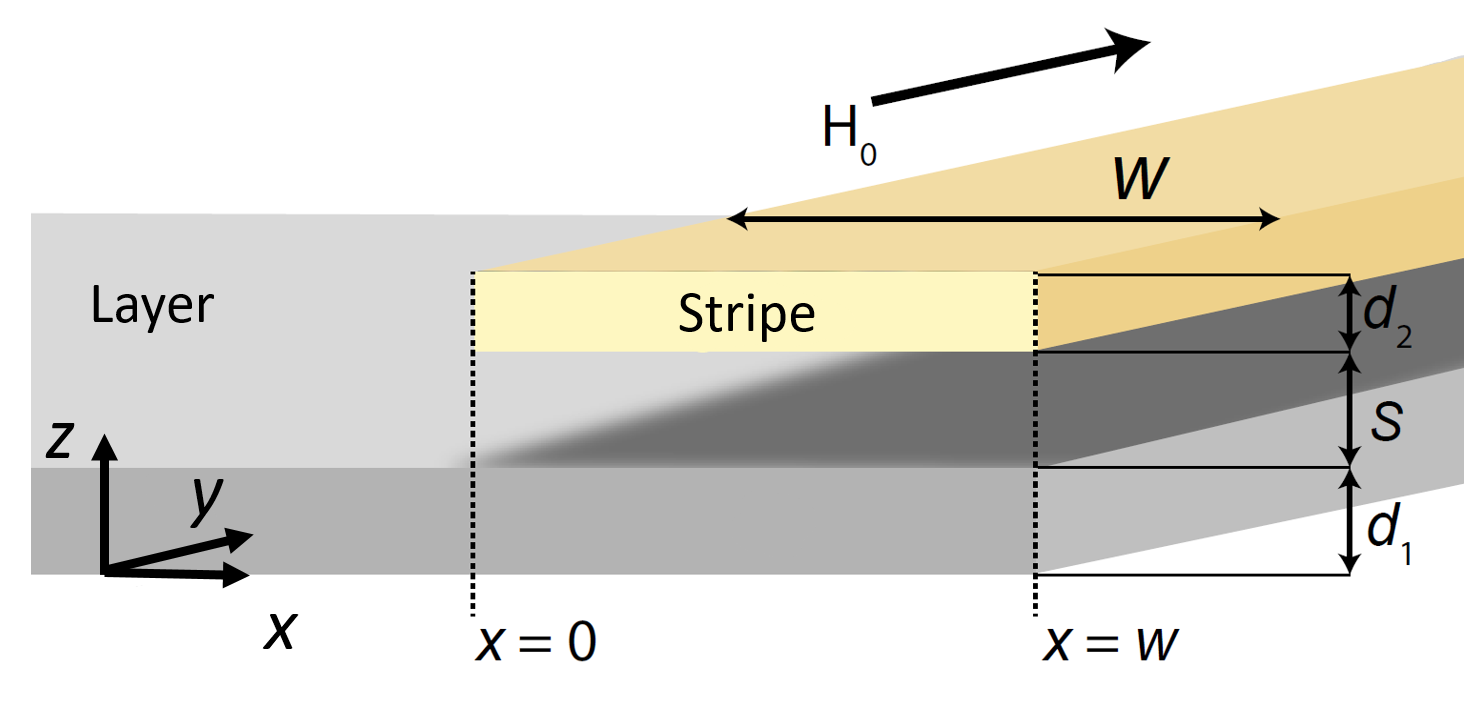}
        \caption{The geometry of the system used in simulations. A Py stripe of thickness~$d_2$ and width~$w$ is separated by a distance $s$ from a semi-infinite Py layer of thickness $d_1$. Both elements have their right edges set at $x=w$. The system is placed in a uniform external magnetic field $H_{0}$ which is parallel to the $y$-axis.}
        \label{fig:geo} 
    \end{figure}
    
\subsection{Description of the system}
    
    
    Let us consider a system presented in Fig.~\ref{fig:geo} that consists of two magnetic elements, a semi-infinite Py layer of thickness $d_1$, and a Py stripe of thickness $d_2$. The separation between the layer and the stripe is described by $s$. Both ferromagnetic elements are infinitely long along the $y$-axis direction and are submerged in a uniform external magnetic field of value $\mu_0 H_0=0.1$~T, which is parallel to the $y$-axis. Material parameters of Py  were chosen as follows: $M_\mathrm{S}=760$~kA/m, $A_\mathrm{ex}=13$~pJ/m, and $\gamma=-176$~radGHz/T. 
    To avoid the influence of the left edge of the system on the reflection of SWs from the right edge with the resonator, in the calculations we considered a Py layer with the width 50 \textmu m, being significantly longer than the wavelength of considered SWs. 


    \begin{figure}[t!]
        \centering
        \includegraphics[width=8.6cm]{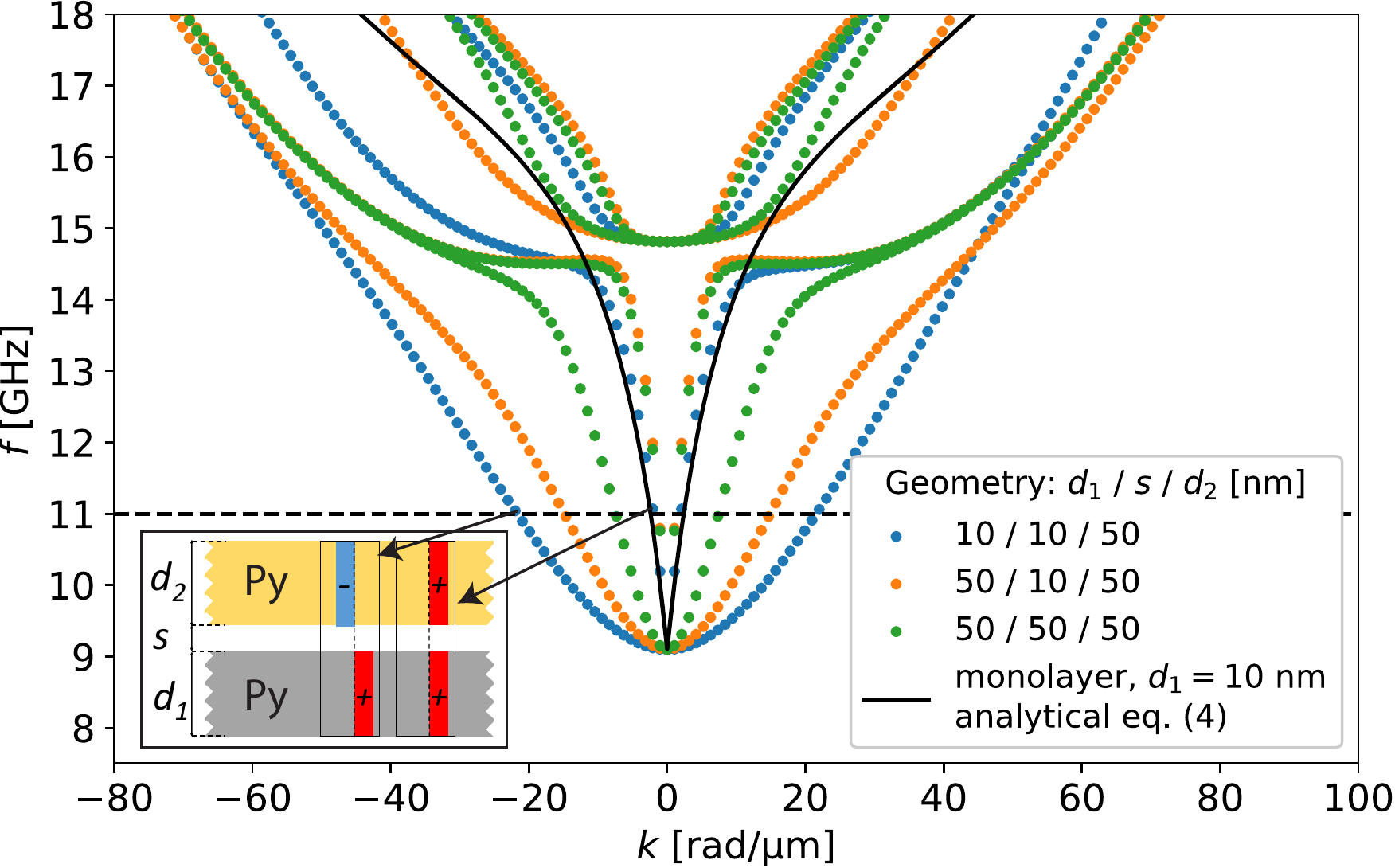}
        \caption{The black solid line represents the dispersion calculated for a layer of thickness $d_1=10$~nm using Eq.~\ref{eq:dispersion}. The blue and orange dots represent dispersion relations calculated numerically for layer $d_1=10$~nm (stripe $d_2=50$~nm, separation $s=10$~nm) and for a layer $d_1=50$~nm (stripe $d_2=50$~nm, separation $s=10$~nm), respectively. The green dots show dispersion calculated for separation $s=50$~nm (in this case $d_1=d_2=50$~nm). In all the cases, dispersions were calculated for infinitely long layers and stripes. Dashed, horizontal line indicates $11$~GHz frequency used in the subsequent calculations. (inset) Sketch of the system used in the dispersions calculations with
        simplified visualization of antisymmetrical short-wavelength (on the left) and symmetrical long-wavelength (on the right) modes.
        }
        \label{fig:dispersions} 
    \end{figure}
    
\subsection{Dispersion relations}
    Before analyzing how the presence of a stripe influences the reflection of SWs, let us examine how the dispersion relation of an infinitely extended bi-layered system changes with respect to the dispersion of a single Py film. The analytically derived dispersion using Eq.~(\ref{eq:dispersion}) for a 10~nm thick Py film is presented in Fig.~\ref{fig:dispersions} by the black-solid line. The FD-FEM computed dispersions for three bi-layered systems with three combinations of Py-layer thicknesses and separation between them are presented by color dots in Fig.~\ref{fig:dispersions}, with the inset explaining the geometry and the symmetry of the modes. For frequencies below 14~GHz, we have two bands instead of one in the case of a single Py film. 
    The band of smaller wavenumber at a selected frequency is related to the symmetric mode of the bi-layer (cf. the schematic representation in the inset), it resembles typical Damon-Eshbach dispersion. Therefore, this band is very similar to the band of a single Py film. The second band is related to the antisymmetric mode for which magnetization oscillates in antiphase in the layers (cf. the schematic representation in the inset), the dispersion of which is parabolic, and has much shorter wavelengths, in comparison to the previous band. 
    
    Comparing the dispersion for different bi-layers shown in Fig.~\ref{fig:dispersions}, we find that in the case of a bi-layer with a smaller value of $d_1=10$~nm, the SWs associated with the short-wavelength band are shorter than in the case of $d_1=50$~nm. One may conclude that, while the separation increases, the wavelengths of the short-wavelength band also increase, since the interaction between both layers decreases, cf. results obtained for $d_1=d_2=50$~nm, $s=10$~nm, and $s=50$~nm in Fig.~\ref{fig:dispersions} represented by orange and green dots, respectively. As expected, in the case of symmetric geometry ($d_1=50$~nm, $s=10$~nm, $d_2=50$~nm) dispersion relation is fully reciprocal, i.e., a mirror symmetry of dispersion with respect to $k=0$ is present. Interestingly, for the geometry $d_1=10$~nm, $s=10$~nm, $d_2=50$~nm a small nonreciprocity is visible (see frequencies above $f=14$~GHz), although the system is composed only of Py. Overall, it means that even in the case of bi-layers composed of the same material, one may easily modify the wavelength of SWs corresponding to short-wavelength bands by changing a separation or the thickness of one of the layers.
    

    The analysis of dispersion relations in single and bi-layered Py films shows that if we consider a system presented in Fig.~\ref{fig:geo}, we have one long-wavelength mode of SWs in a single Py film and two, short- and long-wavelength modes in the bi-layered part. In the following section, we will analyze the effect of the resonance related to the short-wavelength mode on the phase of the reflected SWs for the finite width of the bi-layered part. 
    For these investigations, we select the frequency 11 GHz, which is well below the perpendicular standing SWs frequency. 
    Notably, the wavelength of SW in Py film of thickness 10 nm, 50 nm and 90 nm is $584$~nm, $2651$~nm, and $4724$~nm, respectively. These wavelengths are almost the same as the wavelengths of long-wavelength SW mode in the bi-layer, i.e., $590$~nm for $d_1=10$~nm, $2650$~nm for $d_1=50$~nm and $4600$~nm for $d_1=90$~nm. 

\subsection{Influence of the Py-layer's thickness and bi-layer geometry on the phase of reflected spin waves}    
    
Let us examine how the ferromagnetic layer, stripe, and the non-magnetic spacer thicknesses influence the phase shift dependence on the resonator's width, $w$.
For this purpose, we have performed several FD-FEM computations for geometry presented in Fig.~\ref{fig:geo}. In each simulation, a sweep over the stripe's width $w$ is performed in the range from $2$~nm to $500$~nm with the  $2$~nm step. Thicknesses of the layer and the stripe, as well as separation, were changed separately for each simulation.

\begin{figure}
        \centering
        \includegraphics[width=0.4\textwidth]{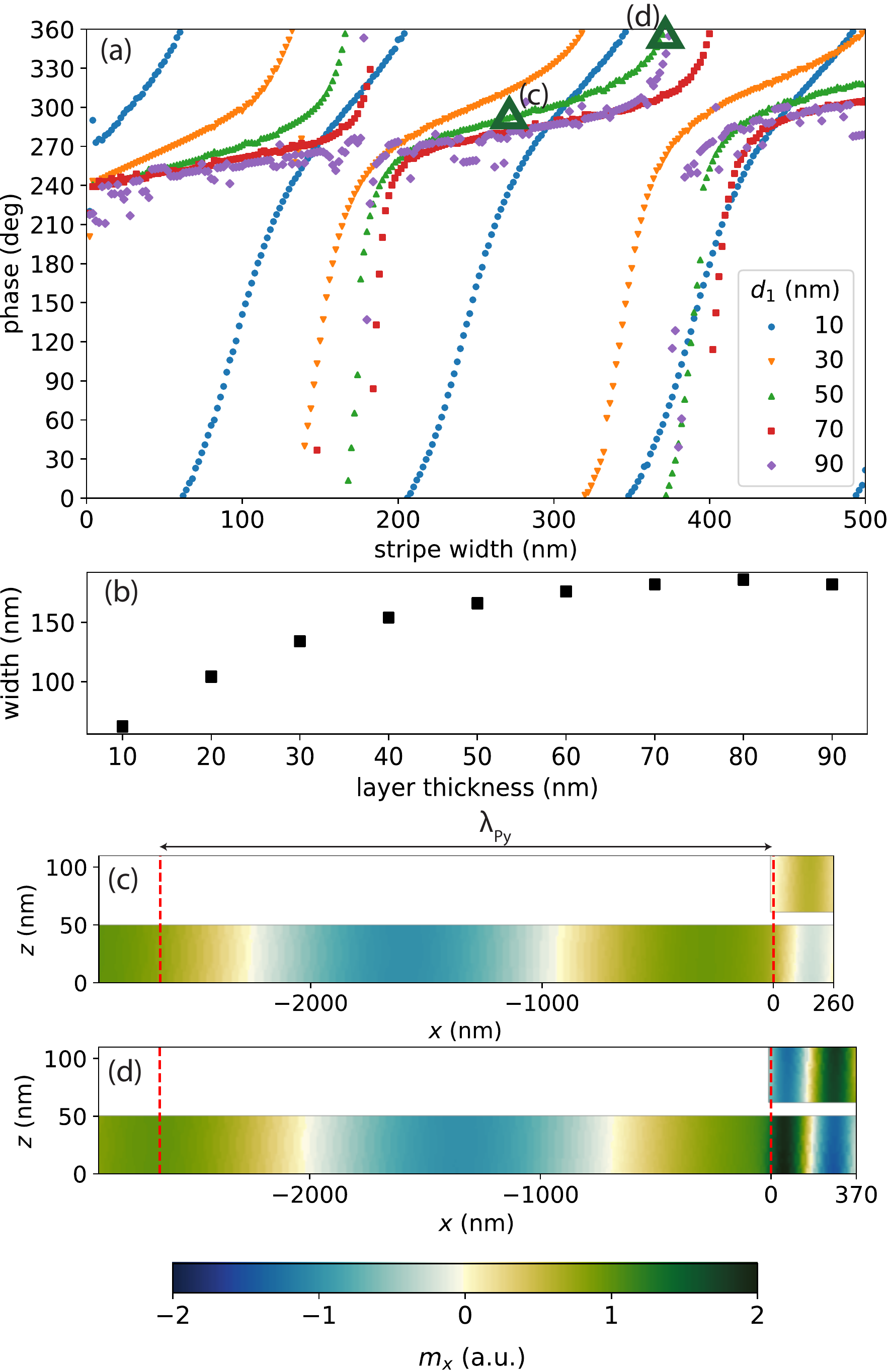}
        \caption{(a) Phase shift as a function of stripe width for layers of different thicknesses at fixed both the separation $s=10$~nm and the stripe thickness $d_2=50$~nm.
        (b) The width of the stripe for which the first resonance occurs as a function of layer thickness. 
        (c) and (d) visualization of SW modes in the system ($d_1=50$ nm, $s=10$ nm, $d_2=50$ nm), (c) stripe $w=260$ nm -- the system is out of resonance, (d) stripe $w=370$ nm -- the system is in the resonance.
        }
        \label{fig:d1_sweep} 
    \end{figure}

Firstly, we analyze how the thickness of the layer $d_1$ influences the phase at constant separation  $s=10$ nm and the thickness of stripe $d_2=50$ nm, see Fig.~\ref{fig:d1_sweep}. In Fig.~\ref{fig:d1_sweep}(a) the phase shift as a function of the stripe's width is shown for five different layer thicknesses equal to 10 nm, 50 nm, 70 nm, and 90 nm. 

Overall, the phase-width dependency has the same features regardless of the layer thickness. Every dependency is characterized by areas of rapid, resonance-like, phase changes, which appear periodically. Those areas are divided by regions of smaller phase change.
In Fig.~\ref{fig:d1_sweep}(d) the visualization of the mode in the resonance area is presented. 
Here, an amplitude of SW below the resonator is noticeably bigger as compared with the rest of the system and an alternating pattern of SW is visible.
We found also an anti-symmetrical character of the mode in the bi-layer (SWs in Py film and stripe oscillate in antiphase). Moreover, the wavelength of SWs in the bilayer is shorter than in the Py film, indicating that the resonance effect is related to the short-wavelength antisymmetric mode.
Fig.~\ref{fig:d1_sweep}(c) presents mode in the region out of resonance, in this case, an increase of the amplitude is absent.

Analyzing how the layer's thickness influences $\varphi(w)$, one may conclude that with increasing thickness of the layer the subsequent resonances appear for greater values of $w$.
This result can be explained by referring to the dispersion relations (cf. Fig.~\ref{fig:dispersions}), which show that with decreasing the layer thickness, the wavelength associated with short SW mode decreases. Therefore, the subsequent resonances appear for shorter $w$.
Additionally, for the thinner layer the areas where phase changes abruptly are wider.
In Fig.~\ref{fig:d1_sweep}(b) the position of the first resonance as a function of the layer's thickness is presented. 
Position of the first resonance at 11 GHz shifts to wider stripes with increasing the layer thickness, starting from $w<70$~nm up to  $w\approx 175$~nm with the increase from $d_1$ from $10$ to $90$~nm, respectively. The stripe's widths with the first resonances are significantly smaller than wavelengths in a single Py layers of checked thicknesses what emphasizes the subwavelength character of the interferometer (cf. dispersion presented in Fig.~\ref{fig:dispersions}). For bigger thicknesses, the dependency becomes too complex to clearly identify the first resonance, due to the quantization of the SWs across the layer thickness.

\begin{figure}
        \centering
        \includegraphics[width=0.4\textwidth]{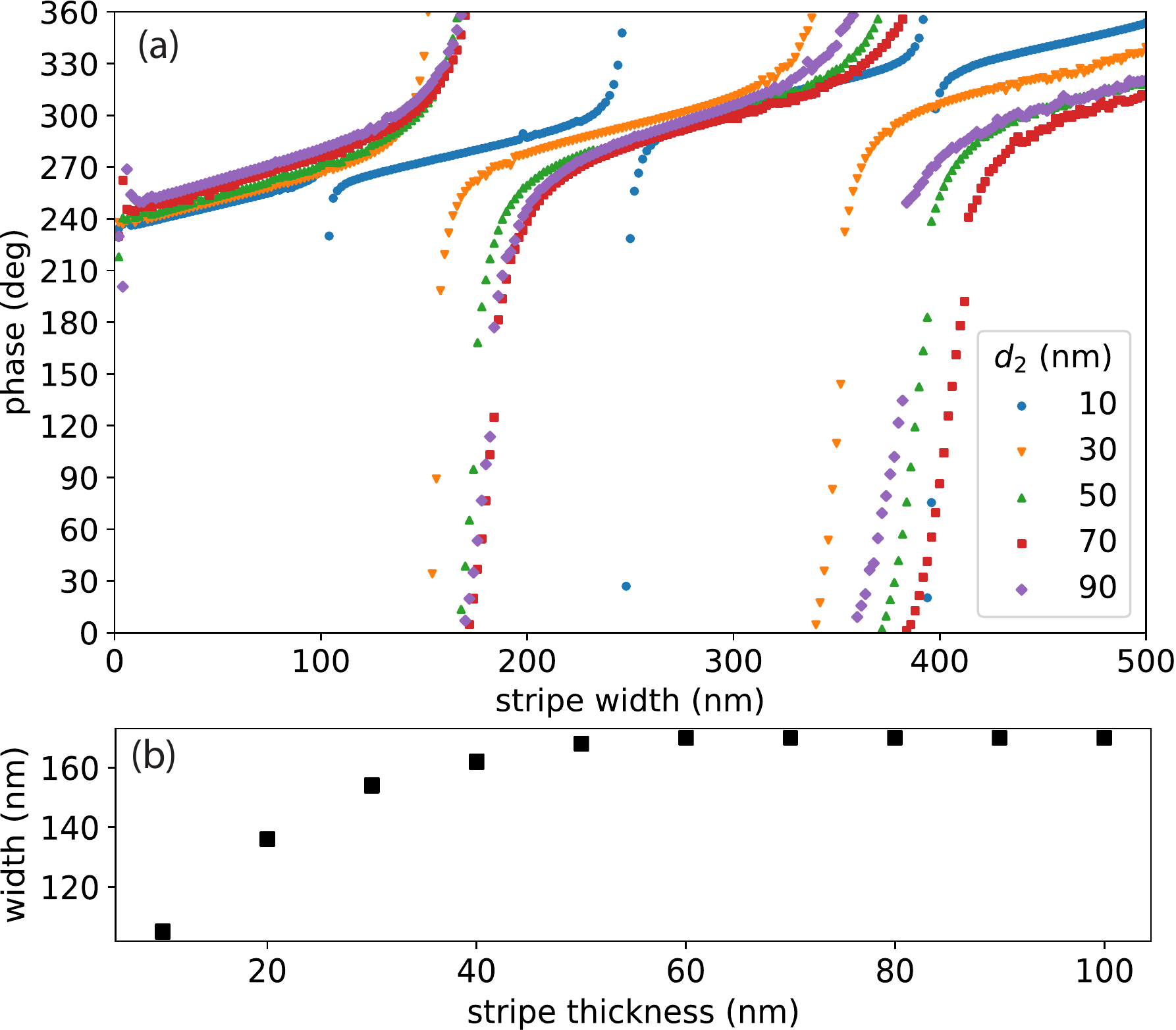}
        \caption{(a) Phase shift as a function of stripe width for the stripes of different thicknesses, with fixed both the separation $s=10$~nm and the layer thickness $d_1=50$~nm.
        (b)~The width of the stripe at which the first resonance occurs as a function of stripe thickness. 
        }
        \label{fig:d2_sweep} 
    \end{figure}

    In Fig.~\ref{fig:d2_sweep} the results of simulations with different stripe thicknesses $d_2$ are shown for fixed $d_1=50$~nm and  $s=10$~nm. Fig.~\ref{fig:d2_sweep}(a) presents the phase of the SW as a function of the stripe width. Here, increasing the thickness of the stripe shifts the position of the first resonance similar to the case presented in Fig.~\ref{fig:d1_sweep}(a). This results from the fact that the thinner one of the layers ($d_1$ or $d_2$) in the bi-layer is, the shorter the SW of the short-wavelength SW mode is (see also Fig.~\ref{fig:dispersions}).
    The main difference between the results in Fig.~\ref{fig:d2_sweep}(a) and Fig.~\ref{fig:d1_sweep}(a) is the fact that the shape of the function $\varphi(w)$, in particular the width of resonances, does not change significantly with changing the thickness of the magnetic element. Resonance areas are separated by plateau areas with the same slope, regardless of the stripe thickness. The position of the first resonance [Fig.~\ref{fig:d2_sweep}(b)] shifts towards wider stripes (from 100 nm) and reaches a plateau ($\approx 170$ nm) for the stripe of thickness $60$~nm, then remains constant for stripes up to $100$~nm, as can be seen in Fig. \ref{fig:d1_sweep}(b). For thicker stripes, a complex phase-width dependency is observed again. 
    
\begin{figure}
        \centering
        \includegraphics[width=0.4\textwidth]{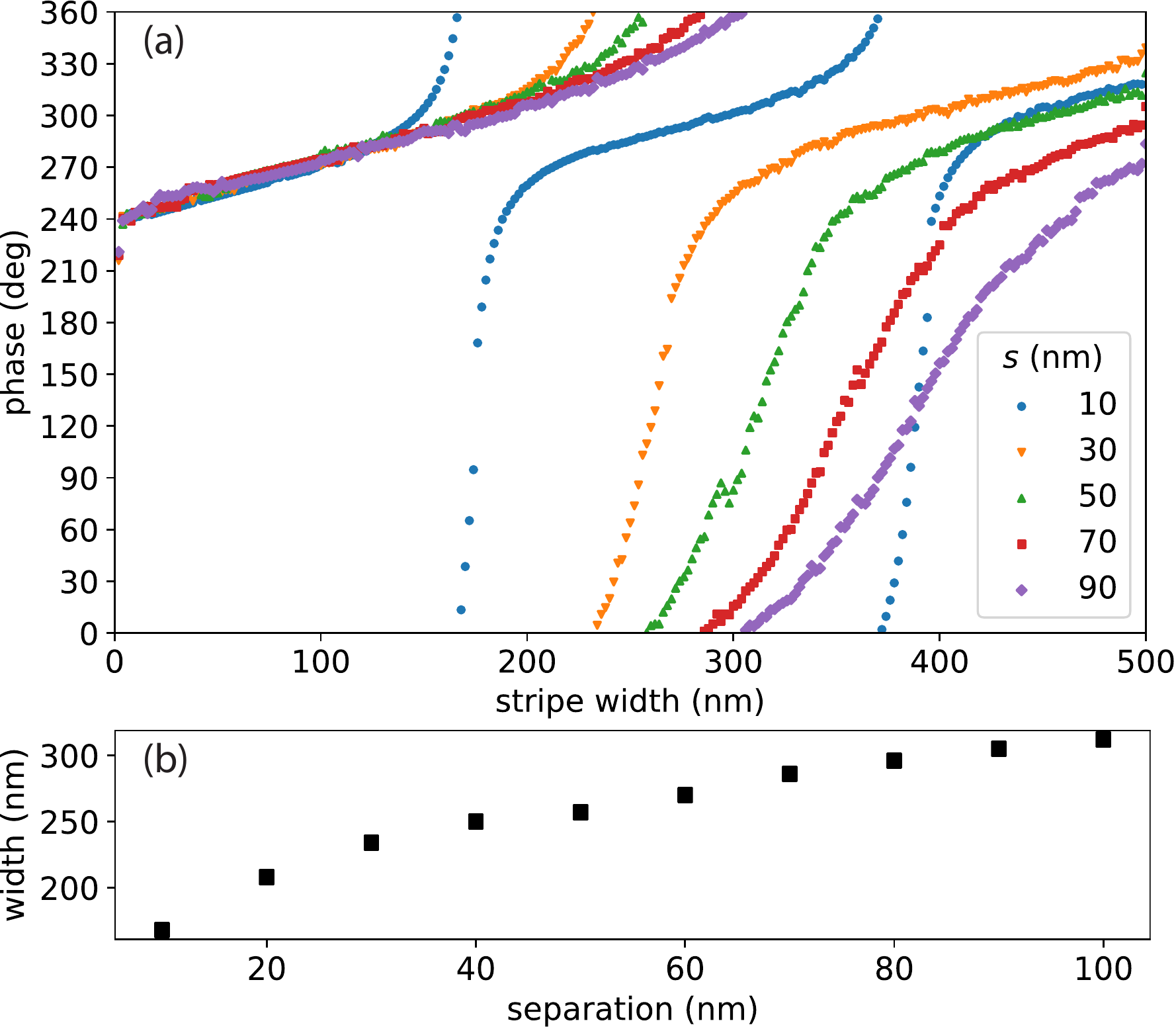}
        \caption{(a) Phase shift as a function of the stripe width for different separation $s$ between the layer and the stripe at $d_1=d_2=50$~nm.
        (b)~The width of the stripe for which the first resonance occurs as a function of 
        separation. 
        }
        \label{fig:s_sweep} 
    \end{figure}

    Fig.~\ref{fig:s_sweep} presents the results of the simulations with different separations between the layer and the stripe at fixed $d_1=d_2=50$~nm. An increase of separation shifts the position of the first resonance to larger widths of the stripe and the shape of dependency evolves as well, as shown in Fig.~\ref{fig:s_sweep}(a). In this case the shape changes in the opposite manner as compared to the results in Fig.~\ref{fig:d1_sweep}(a), namely with the increase of separation the plateau region becomes steeper while in Fig.~\ref{fig:d1_sweep}(a) it becomes flatter. Additionally, as can be seen in Fig.~\ref{fig:s_sweep}(b), the position of the first resonance shifts continuously with the increase of separation, and for maximal considered values of $s$, a maximum shift was not found. 

\section{Conclusions\label{Conclusions}}
We investigate the reflection of spin waves from the edge of a Py film decorated with a narrow Py stripe. We show with FD-FEM numerical simulations that the phase of the reflected spin waves can change by $2\pi$ with dependence on the stripe width. This demonstrates a resonance character which we attribute to the Fabry-Perot resonance of the short-wavelength mode in the bi-layered part of the system. Thus, the proposed system operates as a Gires-Tournois interferometer offering control of the spin-wave phase at sub-wavelength distances, the property important for magnonic applications\cite{Baumgaertl2018,Dobrovolsky2019}. 

We showed that using a single-material-based system, the widths, and the positions of resonances in the magnonic Gires-Tournois interferometer can be controlled over a wide range. In particular, the resonance character depends on the ferromagnetic layer thickness, showing $2\pi$-phase change in a narrow and wide range of the stripe width, for thick and thin films, respectively. The system with the phase steadily varying with the width of the resonator may find application in the design of a metasurface lens for spin waves by proper modulation of the stripe width along its length and so of the reflected spin-wave phase. From the other side, the Gires-Tournois interferometer with a sharp phase change in dependence on the stripe width and its sensitivity for separation between the layers may find application in the design of sensors.


\section*{Acknowledgment}

The research leading to these results has received funding from the Polish National Science Centre projects No. 2019/35/D/ST3/03729, 2018/30/E/ST3/00267. 

\bibliographystyle{IEEEtran}
\bibliography{literature}

\end{document}